\documentclass{article}
\usepackage{spconf,amsmath,graphicx}

\usepackage{multirow}
\usepackage{booktabs}
\usepackage{subcaption}
\usepackage{epstopdf}
\usepackage{url}
\usepackage{cite}

\title{Front-End Diarization for Percussion Separation in Taniavartanam of Carnatic Music Concerts}

\name{%
\begin{tabular}{@{}c@{}}
Nauman Dawalatabad$^{\star}$,
Jilt Sebastian$^{\star}$,
Jom Kuriakose$^{\star}$,\\ 
C. Chandra Sekhar$^{\star}$,
Shrikanth Narayanan$^{\dagger}$, 
Hema A. Murthy$^{\star}$
\end{tabular}}

\address{$^\star$Speech and Music Technology Laboratory, Indian Institute of Technology Madras, India\\
$^\dagger$Signal Analysis and Interpretation Laboratory, University of Southern California, USA}

\begin{document}
\ninept
\maketitle
\begin{abstract}
	Instrument separation in an ensemble is a challenging task.  
    In this work, we address the problem of separating the percussive voices in the taniavartanam segments of Carnatic music. 
    In taniavartanam, a number of percussive instruments play together or in tandem.  
    Separation of instruments in regions where only one percussion is present leads to interference and artifacts at the output, as source separation algorithms assume the presence of multiple percussive voices throughout the audio segment.  
    We prevent this by first subjecting the taniavartanam to diarization.   
    This process results in homogeneous clusters consisting of segments of either a single voice or multiple voices.   
    A cluster of segments with multiple voices is identified using the Gaussian mixture model (GMM), which is then subjected to source separation. 
    A deep recurrent neural network (DRNN) based approach is used to separate the multiple instrument segments.
    The effectiveness of the proposed system is evaluated on a standard Carnatic music dataset. 
    The proposed approach provides close-to-oracle performance for non-overlapping segments and a significant improvement over traditional separation schemes.
\end{abstract}
\begin{keywords}
Diarization, percussion separation, taniavartanam, Carnatic music
\end{keywords}

\section{Introduction}
\label{sec:intro}
Carnatic music is one of the oldest music forms in the world.
It is a classical music form popular in India. 
Carnatic music attributes its origin to Samveda \cite{raghavan1962samaveda}.
Carnatic music has large repositories and is of great interest in music information retrieval (MIR) \cite{serra2014,sadhana2018,venkat2020icassp,Kuri1802:Mridangam,sebastian2017onset,padi2018,venkat2020jasa,jom_akshara}. 
A concert in Carnatic music comprises a lead vocal artist, a violinist, and percussion instrument artists. 
The solo instrument performances attempt to mimic the vocalist.
Mridangam is the lead percussion instrument.
A concert is made up of a sequence of items known as \textit{alapana}, \textit{composition}, and \textit{taniavarnatnam} segments.
The melody of different compositions has different rhythm patterns.
Taniavartanam refers to the solo performance by the percussion artists. 
It is a {\it ``call and response"} section where the percussion artists communicate via different complex phrases.
During taniavartanam, a percussion artist challenges the other percussion artist with a particular complex phrase. 
The other artist gives a fitting response to the challenge. 
 Taniavartanam ends with the overlapped section, where all the percussion artists play certain phrases together.
The instrument diarization system annotates the input taniavartanam with relative instrument labels \cite{dawalatabad18-tani}.
The duration of the phrases used in taniavartanam is dependent on the duration of a cycle,  and a segment can be as short as 500 ms. 
This makes the diarization of taniavartanam a challenging task.

Source separation deals with separating various sources from a mixed audio signal. 
Percussion instrument separation has applications in different research areas as follows; (i) analysis of musical signals to identify the artist \cite{Kuri1802:Mridangam}, (ii) to detect the percussive onsets \cite{sebastian2017onset},  (iii) \textit{Akshara} transcription \cite{jom_akshara}, and (iv) to perform other rhythmic \cite{tian2014study} and melodic analyses \cite{ikemiya2015singing, Jiltrnn}. 
Hence, percussive source separation in taniavartanam is an important problem.
Recently, methods based on deep neural networks for source separation have been used \cite{music_ss_serra_IS19, Jansson2017SingingVS}.
The approaches to source separation are performed on the entire recording \cite{hsu2010improvement,huang2014singing}. 
The approaches in literature treat the whole recording as a mixed signal and attempt to separate the sources in the whole signal.
Recently, authors in  \cite{Goldshtein2018} use diarization and source separation modules in tandem on the speech conversation recording that handles mixtures with more speakers than microphones. 
The percussive voice\footnote{The words \textit{``percussive voice"} and \textit{``voice"} are used interchangeably.} separation system can introduce artifacts in the separated output channels. 
Hence, the output sources from the actual single source regions (i.e., the solo instrument regions) get distorted due to artifacts introduced by the source separating system.

A recording of a taniavartanam can be as long as 5 mins to 30 mins.
The overlapping segments account for  15-20\% of the total duration, and the remaining parts are percussion solo with instrument-homogeneous segments. 
The objective of this work is to avoid performing source separation in the single source regions, thereby retaining the original quality for the single source regions.
This work focuses on two majorly used percussion instruments in taniavartanam; the mridangam and the ghatam.
Our proposed approach is motivated by the ideas in the speech domain \cite{Goldshtein2018}.
To the best of our knowledge, this paper presents for the first time a percussion separation system using diarization as a front-end module.
The proposed system has the following stages. 
\textit{(i) Diarization:} Obtain clusters of solo and mixed audio segments. 
\textit{(ii) Identification:} Obtain the actual identity for each cluster.  
\textit{(iii) Percussion separation:} Perform separation only for mixed audio segments. 
Since the proposed system performs separation only in the mixed audio segments, it retains the original signal quality for non-overlapped regions.

The paper is organized as follows. Section \ref{sec:meth} describes the proposed methodology.
In Section \ref{sec:dataset}, we describe the multi-channel dataset used to demonstrate the effectiveness of the proposed system.
Section \ref{sec:exp} describes the experimental setup, results, and analysis of the proposed system. 
Finally, Section \ref{sec:con} concludes the paper.

\begin{figure}[t]
    \centering
        \includegraphics[scale=0.24]{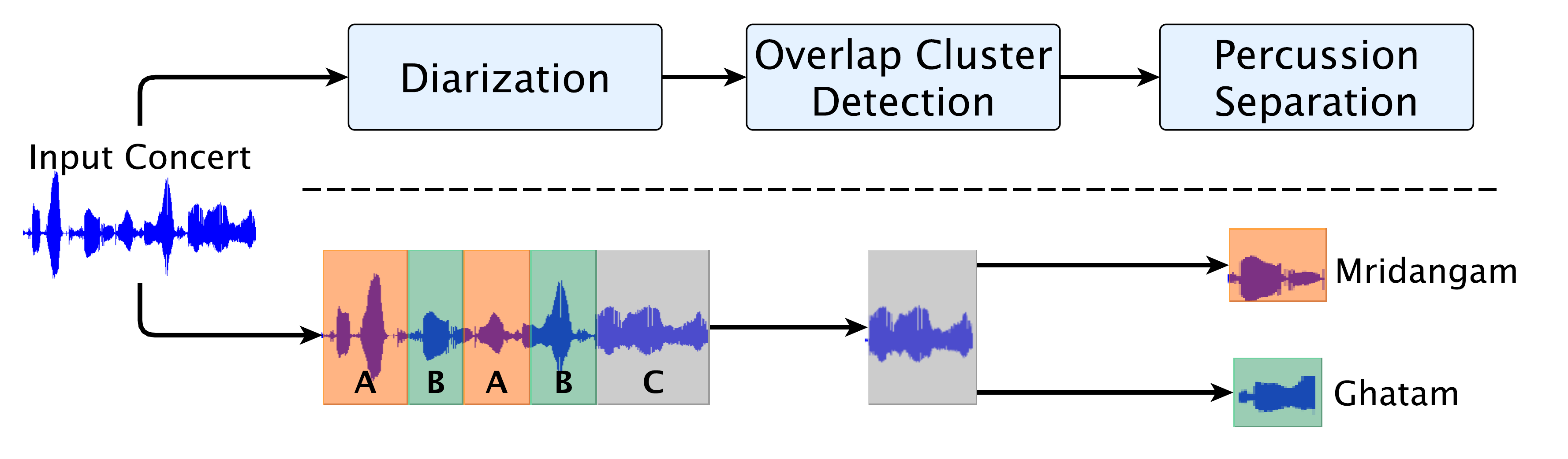}
    \caption{Block diagram of the proposed system.}
    \label{fig:block}
    \vspace{-0.2cm}
\end{figure}

\section{Proposed Methodology}
\label{sec:meth}
\vspace{-0.1cm}
Fig. \ref{fig:block} shows the block diagram of the proposed system.
The stages involved and the individual modules used in the proposed system are explained in detail in the following subsections.

\vspace{-0.1cm}
\subsection{Diarization of Taniavartanam}
\label{sec:diary}
An agglomerative information bottleneck approach to diarization is a bottom-up clustering approach based on information bottleneck (IB) principle \cite{tishby2000-ib}.
It is popular in the speaker diarization literature owing to its low runtime \cite{deepu09-ib,srikanth15-improveRuntime,dawalatabad19:itl,dawalatabad2020novel}.
Recently, we proposed a varying length information bottleneck (VarIB) based diarization system \cite{dawalatabad18-tani}.
This approach has low runtime and has shown good performance in diarizing taniavartanam. 
We diarize the taniavartanam to obtain the clusters for solo and mixed audio segments.
The VarIB approach initializes the clusters by dividing the whole audio recording into short segments such that each segment contains an equal number of strokes.
This distributes instrument information uniformly across the segments.
Let $\mathbf{X_v}$ represent the short segments. 
A GMM is trained using these segments.
The components of the GMM form the relevant variables represented by $\mathbf{Y}$.
The VarIB algorithm arranges segments $\mathbf{X_v}$ into clusters $\mathbf{C}$ using a bottom-up agglomerative approach such that the relevant information $\mathbf{Y}$ is preserved. 
The optimization equation is given as follows,
\begin{equation}
    \mathcal{F} =  \mathrm{I}( \mathbf{Y}, \mathbf{C}) - \mathbf{\frac{1}{\beta}} \mathrm{I}( \mathbf{C}, \mathbf{X_v})
\end{equation}
where $\beta$ is a Lagrange multiplier and $\mathrm{I(\cdot)}$ is the mutual information between the variables. 
The bottom-up clustering is stopped when the threshold on normalized mutual information (NMI) given by \( \frac{\mathrm{I}(\mathbf{Y,C})}{\mathrm{I}(\mathbf{X_v,Y})} \) is reached \cite{deepu09-ib}. 
Finally, the segment boundaries are realigned to obtain the final diarized output \cite{deepu09-klRealign}.

\vspace{-0.1cm}
\subsection{Cluster Identification}
\label{sec:overlap}
We classify the clusters obtained from VarIB based diarization to get their actual identities.
The VarIB system clusters the segments based on the timbre. 
Since the timbre of the overlapped segments differs from the respective solo instruments, the VarIB system forms a  separate cluster for the overlapped segments. 
We propose a two-level classification approach to obtain the actual identities for each of the clusters. 
(i) First, the cluster with overlapping segments is detected based on the cluster likelihood using a GMM trained on separate overlapped data.  
(ii) We then remove the overlapped cluster\footnote{Cluster with overlapping segments is referred to as an overlapped cluster.} and identify the cluster corresponding to ghatam using a GMM trained on separate data consisting of segments from ghatam. 
We do not use the trained mridangam model as it is the most dominating instrument and gets confused easily with the overlapped segments.

It is important to note that the likelihood of the whole cluster is considered during identification instead of segment-wise likelihoods. 
Mridangam is a dominant instrument played for around 60-70\% of the total time (45-50\% in solo, 15-20\% in overlapped). 
It dominates even more in the overlapped sections. 
Hence, it is difficult for a segment-wise classifier to detect the mridangam segments. 
Thus, it is essential to first diarize to get the clusters of audio segments and then classify them to get the actual identities.
These identities are required for obtaining final separated channel outputs. 
Only the segments belonging to the overlapped cluster are fed to the source separation stage.

\vspace{-0.2cm}
\subsection{Percussion Instrument Separation}
\label{sec:ss}
The separation between the percussive sources is very challenging as the time-frequency (T-F) structure shows significant similarity, and the strokes are well-overlapped at certain rhythmic locations. 
We employ a deep recurrent neural network (DRNN) with source-specific mask prediction for performing the separation of two percussion instruments \cite{sebastian2017onset}. 
This framework was originally proposed for singing voice separation \cite{huang2014singing}.
Recently, this framework has shown good performance on separating percussion from musical mixtures in \cite{sebastian2017onset} on the Carnatic music dataset. 
The architecture consists of an input layer,  hidden layers with recurrent connections, and two output layers corresponding to the mridangam and the ghatam. Magnitude spectrogram is used as the feature, and the estimated mask is mixed with a mixture spectrogram for obtaining the magnitude spectrogram estimates of the sources. 
This is then mixed with the noisy phase to generate the individual channels.

The loss function is regularised not only to minimize the difference between the estimated and the corresponding target percussive voices but also to maximize the difference between the estimated and the other percussive voice, owing to a very small timbre difference. We use a single model to learn masks for both of the sources. The objective function for this discriminative training is given by,
\begin{equation}
\label{def:rnn}
||\widehat{y}_{m}-y_{m}||^2 +||\widehat{y}_{g}-y_{g}||^2- \gamma(||\widehat{y}_{m}-y_{g}||^2 + ||\widehat{y}_{g}-y_{m}||^2)
\end{equation}
where $\widehat{y}$ and $y$ are the estimated and the original magnitude spectra, and $m$ and $g$ denote mridangam and ghatam, respectively. We want the small timbre difference to be captured. Hence, the discriminative term is added to the loss function. The magnitude spectrogram prediction is formulated as a soft mask estimation problem where each T-F bin is classified as either percussive or non-percussive voice with some probability. 
After percussion separation, an instrument channel is formed in the non-overlapping part by concatenating the segments of the instrument when it is present, and by replacing the temporal positions by zeros when it is absent.

The proposed system is modular; hence each module can be replaced by other suitable algorithms.
In this paper, we use modules that have shown good performance on the Carnatic music dataset.

\section{Dataset description}
\vspace{-0.2cm}
\label{sec:dataset}
We use live multi-channel audio recordings of the Carnatic music concerts from the standard  \textit{MusicBrainz} \cite{tani-data} repository developed as a part of CompMusic Project\footnote{CompMusic [accessed 04-Mar-2021]: http://compmusic.upf.edu/}. Six live concert recordings are chosen randomly that have mridangam and ghatam played in the taniavartanam section. 
This section is manually extracted, and the dataset is created by annotating the instrument change points and the overlapped sections.
 For each of the taniavartanam, the channels corresponding to mridangam and ghatam are mixed to form one audio in the test phase. 
 The details of the dataset are given in Table \ref{tab:dataset}.
We use one concert (AK) to train the source separation model, and the rest are used for testing purpose.

\begin{table}[t]
\centering
\caption{Details of the dataset. Durations of Mridangam (Mri), Ghatam (Ghat), and overlapped part in each of the recordings in the dataset are mentioned.}
\vspace{-0.25cm}
\resizebox{0.37\textwidth}{!}{%
\begin{tabular}{ccccc}
\toprule
\multirow{2}{*}{Concert ID} & \multicolumn{3}{c}{Segments (mm:ss)} & \multirow{2}{*}{\begin{tabular}[c]{@{}c@{}}Total Length\\ (mm:ss)\end{tabular}} \\ \cmidrule(l){2-4}
                         & Mri        & Ghat      & Overlap     &                                      \\  \midrule
AK                       & 06:13      & 03:52     & 01:36       & 11:41                                                                           \\
DR                       & 08:29      & 04:26     & 02:09       & 15:04                                                                           \\
MD                       & 03:13      & 02:06     & 01:28       & 06:47                                                                           \\
RM                       & 06:22      & 05:16     & 01:57       & 13:35                                                                           \\
RT                       & 03:47      & 03:28     & 01:47       & 09:02                                                                           \\
SR                       & 09:18      & 07:47     & 04:47       & 21:52                                                                           \\ \midrule
Total                    & 37:22      & 26:55     & 13:44       & 01:18:01                                                          \\ \bottomrule             
\end{tabular}
}
\label{tab:dataset}
\end{table}

We use the Tani-Dev dataset that was introduced in \cite{dawalatabad18-tani} to tune the hyperparameters for diarization systems.
This dataset contains ten taniavartanam mono-channel recordings, which are also a part of the standard \textit{MusicBrainz} repository.
The models for the cluster identification is also trained on this dataset. 
All recordings are sampled at 44.1 kHz.

\section{Experimental Studies}
\label{sec:exp}
In this section, we demonstrate the effectiveness of the proposed approach on standard Carnatic datasets.

\vspace{-0.2cm}
\subsection{Experimental Setup}
\vspace{-0.1cm}
The input to the diarization module is 19-dimensional MFCC features. 
The threshold on NMI and $\beta$ for both IB\footnote{Code: https://github.com/idiap/IBDiarization} \cite{deepu09-ib}  and VarIB \cite{dawalatabad18-tani} systems were set to 0.4 and 10, respectively. 
The fixed duration segment length for the IB system was set to 2 seconds, and the minimum number of strokes for the VarIB system was set to 15.
As there are two percussion instruments, the maximum number of clusters for IB and VarIB was set to three. 
These hyperparameters are also the same as used in \cite{dawalatabad18-tani}.
The number of GMM components for the cluster identification was set to three. 
For the source separation module, a 1024 point short-time Fourier transform is used as the feature. 
The DRNN architecture\footnote{Code: https://github.com/posenhuang/deeplearningsourceseparation} has 500 nodes at three hidden layers with recurrent connections (stacked RNN) and 513 nodes at the output layer. 
ReLU non-linearity is applied at the output of each layer.
Mean squared error is used as the loss function, and the hyper-parameter $\gamma$ is set at 0.08. 
All the hyperparameters were tuned to obtain the best results on the development dataset.

Percussion instrument separation on the entire taniavartanam segment without diarization is considered as the traditional system. 
The final output of the proposed system consists of the concatenation of separated channels from the overlapped clusters and corresponding segments from the non-overlapped clusters.
Note that all these segments are obtained from diarization. 
Recordings of the individual channels form the oracle system for non-overlapped segments.

\vspace{-0.1cm}
\subsection{Results and Discussion}
\label{ssec:eval}
We not only report the final separation performance  but also report the intermediate performances of each module in the proposed system.
The results are discussed in the following subsections.

\begin{table}[t]
\centering
\caption{Diarization Error Rate (DER) for IB and VarIB systems on different datasets.}
\vspace{-0.25cm}
\resizebox{0.45\textwidth}{!}{%
\begin{tabular}{ccccccccc}
\toprule
\multirow{2}{*}{Sys.} & \multirow{2}{*}{\begin{tabular}[c]{@{}c@{}}Dev Set \end{tabular}} & \multicolumn{7}{c}{Test Set}                     \\ \cmidrule{3-9}
                     &                                                                               & AK  & DR   & MD   & RM  & RT  & SR   & Avg. \\ \midrule
IB                   & 17.9                                                                          & 7.2 & 33.4 & 30.1 & 5.4 & 3.8 & 29.3 & 20.0      \\
VarIB                & 13                                                                            & 6.9 & 13.5 & 18.9 & 6.5 & 3.8 & 13.3 & 10.6     \\
\bottomrule
\end{tabular}}
\label{tab:der}
\end{table}

\vspace{-0.3cm}
\subsubsection{Diarization Module}
\vspace{-0.1cm}
 We use a standard tool from NIST\footnote{https://github.com/nryant/dscore/blob/master/scorelib/md-eval-22.pl} to evaluate the diarization error rate (DER).
DER is the sum of missed segments, false alarms, and instrument errors. 
Since the silence is negligible in the taniavartanam, voice activity detection is not needed. 
Hence, the missed segments and false alarms are almost zero. 
The lower the DER, the better is the clustering solution. 
We use a strict forgiveness collar of 0.15 sec in evaluating DER \cite{dawalatabad18-tani}. 
The forgiveness collar is essential as it ignores both, the errors made by the system and also the human errors made during ground truth annotations within a specified window.

Table \ref{tab:der} shows the DER obtained on development and test datasets.
The DER reported for the development set is averaged over all the ten recordings, while for the test set, we report individual DER for each recording.
It can be seen that the VarIB system outperforms the traditional IB system for most recordings.
The average DER for VarIB is absolute 9.4\% (47\% relative) better than the baseline system on test data.
This also confirms that varying length segmentation yields a significantly better clustering solution \cite{dawalatabad18-tani}.

\begin{table}[t]
\caption{Average cluster purity and identification accuracy (Acc.) for VarIB diarization followed by cluster identification.}
\vspace{-0.25cm}
\centering
\resizebox{0.45\textwidth}{!}{%
\begin{tabular}{ccccccccc}
\toprule
\multirow{2}{*}{Metric} & \multirow{2}{*}{Dev Set} & \multicolumn{7}{c}{Test Set}                   \\ \cmidrule{3-9}
                        &                      & AK   & DR   & MD   & RM   & RT   & SR   & Avg.  \\ \midrule
Purity                  & 0.89                 & 0.91 & 0.86 & 0.80 & 0.93 & 0.95 & 0.85 & 0.88 \\
Acc. (\%)                & 86.0                 & 90.7 & 85.5 & 79.8 & 92.7 & 95.3 & 85.2 & 88.2 \\
\bottomrule
\end{tabular}}
\vspace{-0.25cm}
\label{tab:acc}
\end{table}

\vspace{-0.1cm}
\subsubsection{Cluster Identification}
\vspace{-0.1cm}
 The performance of the system after VarIB based diarization and cluster identification is measured in terms of averaged cluster purity and identification accuracy.
Since the segments in a cluster can be partially overlapped/pure, the purity and accuracy are reported in terms of time duration using an open-source tool \cite{pyannote}.
Purity is calculated as the ratio of the sum of the duration of the most frequent segments in a cluster to the duration of that cluster. 
Hence, a cluster with purity 1.0 is a perfectly pure cluster.
The accuracy is calculated in terms of percentage of the time duration of the correctly identified taniavartanam.

Table \ref{tab:acc} shows the average cluster purity and accuracy for each concert for the cluster identification stage. 
The observed average cluster purity is 0.89 and 0.88 on development and test data, respectively. 
The  GMM classifier shows a good average accuracy of 86.0\% and 88.2\% on the train and test data, respectively.
This shows that the proposed GMM classifier detects all the clusters obtained from the diarization module correctly. 
However, the reported frame-level accuracy is not 100\% because of a few short segments that get wrongly clustered during the diarization stage.
The exact value of these errors can be obtained as 1-purity from Table \ref{tab:acc}.

\vspace{-0.1cm}
\subsubsection{Percussion Separation}
\vspace{-0.1cm}
\label{ssec:separation}
 We use blind source separation evaluation (BSS Eval) metric \cite{vincent2006performance} to measure the quality of the source separation. 
 The results on the length-weighted signal-to-distortion ratio (SDR) averaged across the segments (Global SDR) are reported. 
 This overall measure accounts for both the artifacts introduced and the interfering source present in the separated channel. 
SDR considers the overall quality, which includes the other BSS measures, namely, SAR (Signal to Artifact Ratio) and SIR (Signal to Interference Ratio).
Hence, these measures are not reported separately in the paper.
This evaluation measure requires the knowledge of original sources for computing the separation quality. 
Since both the channels are active for the overlapped audio, we can measure the SDR value. 
The oracle system consists of source separation on the ground truth segments, and the proposed system contains the same separation on the segments with boundaries coming from the diarization system.

\begin{table}[t]
\caption{Global SDR for overlapped segments of Proposed (diarized segments) and Oracle (ground truth segments) systems. The GSDR for the traditional system is same as oracle for overlapped segments.}
\vspace{-0.25cm}
\centering
\resizebox{0.43\textwidth}{!}{%
\begin{tabular}{cccccccc}
\toprule
\multirow{2}{*}{System}  & \multicolumn{6}{c}{Test Set}                     \\ \cmidrule{2-7}                                                                                              & DR   & MD   & RM  & RT  & SR   & Avg. \\ \midrule
\multirow{1}{*}{Proposed }                                                                                  & 1.475 & 0.914 & 3.953 & 4.037 & 3.867 & 2.849    \\ 
\multirow{1}{*}{Oracle}                                                                                  & 1.138 & 0.896 & 4.076 & 4.032 & 3.406 & 2.710    \\ 
\bottomrule
\end{tabular}}
\label{tab:ss1}
\end{table}

The GSDR of the proposed approach and the oracle system for overlapped segments are shown in Table \ref{tab:ss1}. 
Since the traditional system blindly separates the whole recording, the GSDR values are the same as the oracle system for overlapped segments.
The GSDR values are inferior to other musical separation tasks as these are live concert recordings, and there is a high similarity between the T-F structures of the percussion instruments considered. Moreover, the timbre for the same instrument varies across the concerts (based on the base frequency, i.e., tonic). 
Note that the proposed method has better GSDR values than the oracle system for some test sets. 
This is because few overlapping parts are classified as non-overlapping, and for these segments, the GSDR is lesser than the oracle average. As they are not included in the GSDR computation only for the proposed system, the resulting measure will be higher than the oracle.

We have to evaluate the system for non-overlapped segments separately. 
This is because though the ground truth should ideally have zero amplitudes in the inactive channel,  it has traces of other channel and microphone noises. 
The input mixture is the sum of this inactive channel and the instrument channel. Providing a zero amplitude signal or a small noisy signal (the estimated inactive output) causes problems for calculating BSS evaluation metrics as they are computed by projecting the estimated source energy to the original source sub-spaces \cite{vincent2006performance}. Nevertheless, the GSDR can be computed for the traditional system since both channels are active for non-overlapped segments (reported later in this section).

\begin{figure}[t]
    \centering
    \includegraphics[scale=0.45]{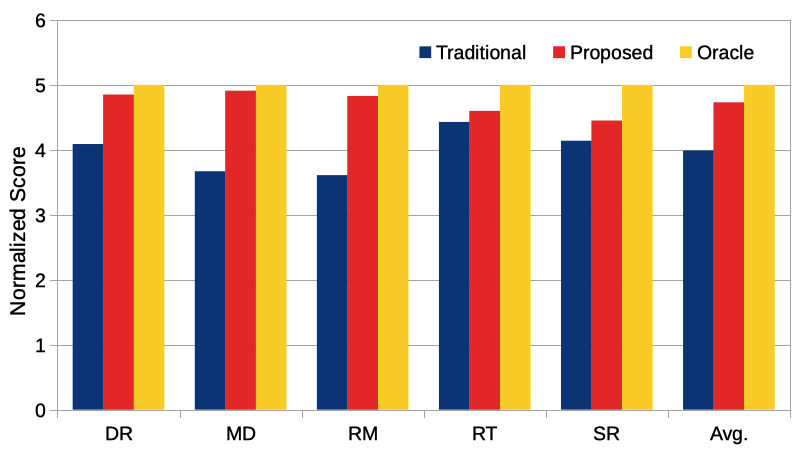}  
    \vspace{-0.2cm}
    \caption{Subjective evaluation scores on non-overlapped segments.} 
    \label{fig:subjective}
   \vspace{-0.2cm}
\end{figure}

For the non-overlapping segments, we perform qualitative analysis to compare the performances of (i) the oracle system (ground truth segments and no separation), (ii) the proposed system (diarization-based segments and no separation), and (iii) the traditional system (ground truth segments with separation).
Subjective mean opinion score (MOS) \cite{mos2016} test is performed with 15 subjects listening to four randomly selected examples (2 from mridangam and ghatam each) from non-overlapped clusters in each of the concerts. 
For each example, having a length of 20 sec, three systems are evaluated after hiding their identity. 
All the examples were randomly shuffled before presenting them to the subject.  
 The subject was asked to listen to the audio clips from all the three different systems and then rate them between 1 to 5 (5 being the highest) based on the perceived quality of audio.
 The final evaluation scores averaged over all subjects are given in Fig. \ref{fig:subjective}.

The average GSDR value of the traditional system on non-overlapping segments is 2.727 (DR=1.776, MD=1.708, RM=3.855, RT=3.432, SR=2.862).
GSDR for non-overlapped clusters is very high (ideally infinity) for the proposed approach as the output signal is similar to the respective individual channels. 
This is also observed for most of the concerts in the subjective evaluation in Fig. \ref{fig:subjective}. 
It can be seen from the Fig. \ref{fig:subjective} that the scores of the proposed system are close to that of the oracle system.
The small average reduction  in the evaluation score for the proposed approach with respect to the oracle system (0.27) is mainly due to the
noise captured in the inactive channel, which makes the sum of two channels different from the actual voices (active channel).
Hence, a significant improvement in performance over a traditional separation system is achieved for the non-overlapping segments by incorporating diarization and subsequent cluster identification.

\begin{figure}[]
    \centering
    \includegraphics[scale=0.23]{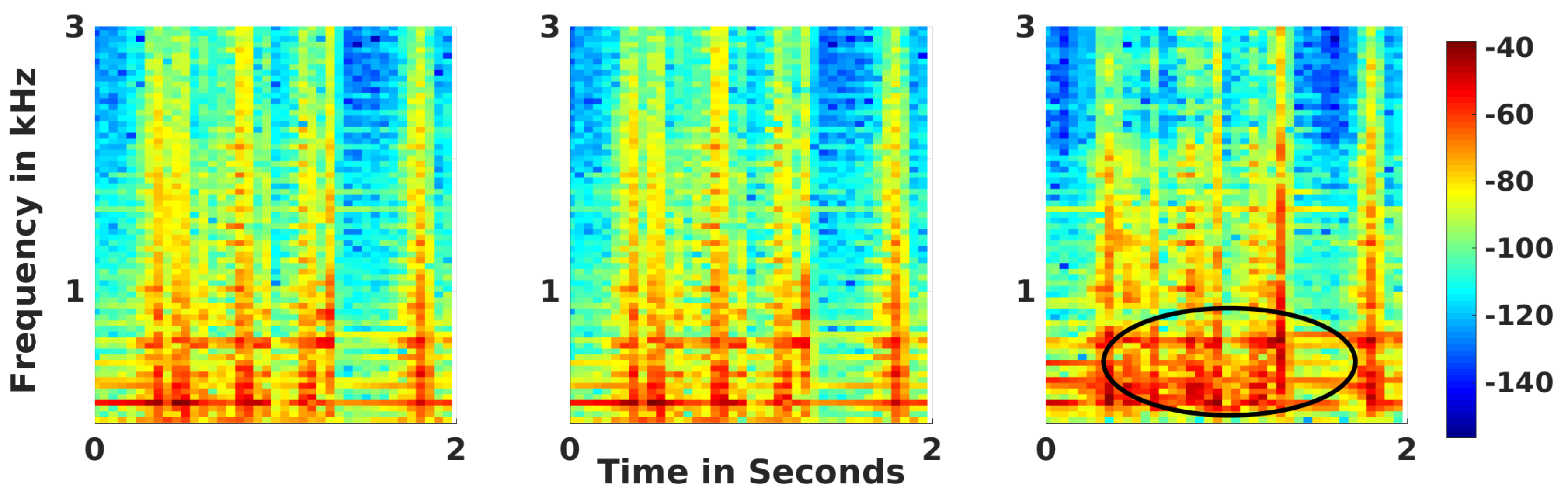}
    \vspace{-0.2cm}
    \caption{An example excerpt from non-overlapped mridangam segment. Outputs of the proposed (\textit{left}),  ground truth (\textit{middle}), and traditional separation (\textit{right}) systems. 
    Observe that the artifacts are introduced in the output of traditional system.
    }
    \label{fig:spectro}
   \vspace{-0.3cm}
\end{figure}

To observe the system artifact, we also plot the spectrogram for a non-overlapped segment for all the three systems, as shown in Fig. \ref{fig:spectro}.
It can be observed clearly in Fig. \ref{fig:spectro} that the artifacts are introduced by the traditional system in the separated segment. 
Also, the traditional source separation system would consider the whole non-overlapped section and blindly separate it, generating the percussive voices for the inactive channel as well. 
The proposed system does not separate the non-overlapped segments. 
The performance gain achieved by the proposed approach is huge since it is improper to separate a non-overlapping segment.

\section{Conclusion} 
\label{sec:con}
We propose a diarization-driven percussion separation pipeline for taniavartanam.
The advantage of this approach is that the systems in each of the modules can be easily replaced with another algorithm.
The proposed system is able to reduce the artifacts arising in the traditional separation systems owing to the separation performed in the non-overlapping segments.
The performance of the proposed system on the non-overlapping segments is close to oracle.
This suggests the importance of diarization as a front-end to reduce the deterioration effects (artifacts and interferences) of traditional source separation systems.

\bibliographystyle{IEEEbib}
\bibliography{references}

\end{document}